\documentclass[manuscript]{aastex}

\newcommand{\Gaia}{{\it Gaia}}        
\newcommand{\Hipp}{{\it Hipparcos}}        

\newcommand{\HST}{{\it HST}}

\def\pmb#1{\setbox0=\hbox{#1}
  \kern-.02em\copy0\kern-\wd0
  \kern.01em\copy0\kern-\wd0
  \kern.01em\copy0\kern-\wd0
  \kern.01em\copy0\kern-\wd0
  \kern.01em\copy0\kern-\wd0
  \kern-.02em\raise.01em\box0 }

\def\ref#1#2{$^{#1}$}

\shorttitle{Parallax of Polaris B}
\shortauthors{Bond et al.}

\begin{document}

\title{{\em Hubble Space Telescope\/} Trigonometric Parallax of
Polaris~B, Companion of the Nearest Cepheid\altaffilmark{*}}

\author{
Howard E. Bond\altaffilmark{1,2,3},
Edmund P. Nelan\altaffilmark{2},
Nancy Remage Evans\altaffilmark{4},
Gail H. Schaefer\altaffilmark{5},
and
Dianne Harmer\altaffilmark{6}
}

\altaffiltext{*}
{Based in part on observations made with the NASA/ESA {\it Hubble Space
Telescope}, obtained by the Space Telescope Science Institute. STScI is operated
by the Association of Universities for Research in Astronomy, Inc., under NASA
contract NAS5-26555.}

\altaffiltext{1}
{Department of Astronomy \& Astrophysics, Pennsylvania State University,
University Park, PA 16802, USA; heb11@psu.edu}

\altaffiltext{2}
{Space Telescope Science Institute, 
3700 San Martin Dr.,
Baltimore, MD 21218, USA;
nelan@stsci.edu}

\altaffiltext{3}
{Visiting astronomer, Kitt Peak National Observatory, National Optical
Astronomy Observatory, which is operated by the Association of Universities for
Research in Astronomy (AURA) under a cooperative agreement with the National
Science Foundation.  
}

\altaffiltext{4}
{Smithsonian Astrophysical Observatory, MS 4, 60 Garden St., Cambridge, MA
02138, USA}

\altaffiltext{5}
{The CHARA Array of Georgia State University, Mount Wilson Observatory, Mount
Wilson, CA 91023, USA; schaefer@chara-array.org}


\altaffiltext{6}  
{National Optical Astronomy Observatory, 950 North Cherry Avenue, Tucson, AZ
85726, USA}

\begin{abstract}

Polaris, the nearest and brightest Cepheid, is a potential anchor point for the
Leavitt period-luminosity relation. However, its distance is a matter of
contention, with recent advocacy for a parallax of $\sim$10~mas, in contrast
with the \Hipp\/ measurement of $7.54\pm0.11$~mas. We report an independent
trigonometric parallax determination, using the Fine Guidance Sensors (FGS) on
the {\it Hubble Space Telescope}. Polaris itself is too bright for FGS, so we
measured its 8th-magnitude companion Polaris~B, relative to a network of
background reference stars. We converted the FGS relative parallax to absolute,
using estimated distances to the reference stars from ground-based photometry
and spectral classification. Our result, $6.26\pm 0.24$~mas, is even smaller
than found by \Hipp. We note other objects for which \Hipp\/ appears to have
overestimated parallaxes, including the well-established case of the Pleiades.
We consider possible sources of systematic error in the FGS parallax, but find
no evidence they are significant. If our ``long'' distance is correct, the high
luminosity of Polaris indicates that it is pulsating in the second overtone of
its fundamental mode. Our results raise several puzzles, including a long
pulsation period for Polaris compared to second-overtone pulsators in the
Magellanic Clouds, and a conflict between the isochrone age of Polaris~B
($\sim$2.1~Gyr) and the much younger age of Polaris~A. We discuss possibilities
that B is not a physical companion of A, in spite of the strong evidence that it
is, or that one of the stars is a merger remnant. These issues may be resolved
when \Gaia\/ provides parallaxes for both stars.

\end{abstract}

\keywords{astrometry --- stars: distances ---  stars: evolution --- stars:
individual (Polaris) --- stars: Cepheids}

\section{Polaris: Nearest Cepheid, Controversial Distance}

Cepheid variables are among the primary distance indicators for the
extragalactic distance scale, and they provide critical tests of
stellar-evolution theory. The North Star, Polaris ($\alpha$~Ursae Minoris), is
of special interest as the nearest and brightest Cepheid. It has a relatively
short pulsation period of 3.969~days (Fernie et al.\ 1995). Because it is so
nearby, Polaris can potentially serve as one of the anchor points near the
short-period end for zero-point calibration of the Leavitt period-luminosity
relation (e.g., Feast \& Catchpole 1997; van Leeuwen et al.\ 2007). However, it
is necessary to consider several peculiarities presented by Polaris, as
described below. In particular, it is crucial to determine whether it pulsates
in the fundamental mode or an overtone.

{\bf The majority of classical Cepheids have characteristic asymmetric light
curves with high amplitudes. It was recognized many years ago---for a historical
review see Beaulieu et al.\ (1995)---that there is a separate class of Cepheids
with low pulsation amplitudes and nearly symmetric sinusoidal light curves.
These objects are interpreted as Cepheids pulsating in the first overtone of the
fundamental period. They can be recognized from examination of the light curves,
or more rigorously by calculating a Fourier decomposition (e.g., Simon \& Lee
1981; Poretti 1994).  However, this approach has been difficult for Polaris
itself, because its pulsation amplitude is very small, making the Fourier
coefficients difficult to determine. If the distance is known, overtone Cepheids
can also be distinguished because of luminosities lying above the classical
Cepheids in the Leavitt relation. }

Polaris belongs to a triple system. Its well-known visual companion, Polaris~B,
is an 8th-mag F3~V star, lying $18''$ from the Cepheid. The Cepheid itself was
known for many years to be a single-lined spectroscopic binary with a period of
$\sim$30~yr (Roemer 1965; Kamper 1996, hereafter K96,  and references therein).
But the close companion remained unseen until it was finally detected directly
in {\it Hubble Space Telescope\/} (\HST\/) ultraviolet images by members of our
team (Evans et al.\ 2008), at a separation of $0\farcs17$ from Polaris~A\null.
We inferred a spectral type of F6~V for the close companion, designated
Polaris~Ab, based on its UV brightness and dynamical mass.

The extensive literature on Polaris was reviewed by K96 and Wielen et al.\
(2000), both of whom presented strong evidence that Polaris~B is a true physical
companion of the Cepheid. The physical association with A is supported by
agreement in radial velocity, proper motion, metallicity, and approximate
distance estimates based on photometry and the spectral type of Polaris~B\null.
An examination of the position angle and angular separation of Polaris~B
relative to A (K96; Evans et al.\ 2008) showed that the position angle has
remained essentially constant for more than the past two centuries, and there
has been a slow reduction in the separation at a rate of $-1.67\pm0.19\,\rm
mas\, yr^{-1}$ (Evans et al.\ 2008), consistent with orbital motion at a period
of order $10^5$~yr. Since the absolute proper motion of Polaris~A is about
$46\,\rm mas\,yr^{-1}$, the angular tangential motions of A and B agree to
within $\sim$4\%. Usenko \& Klochkova (2008) provided additional evidence that
the radial velocities of A and B are very similar. A more recent, extensive
review of our knowledge of the Polaris system is given by Turner (2009).

The distance of Polaris has been controversial. Historical ground-based
photographic trigonometric parallaxes of Polaris have such large uncertainties
that they are of limited utility. The Yale catalog (van Altena et al.\ 1995)
gives an average parallax of $4.0\pm3.3$~mas from several determinations.
However, Turner et al.\ (2013, hereafter TKUG13) argue that magnitude-dependent
corrections to the Allegheny Observatory parallaxes would increase the
ground-based value to $11\pm4$~mas.

The \Hipp\/ astrometric mission yielded an absolute parallax for Polaris~A of
$7.56\pm0.48$~mas (ESA 1997), modified to $7.54\pm0.11$~mas in the re-reduction
by van Leeuwen (2007), corresponding to a distance of $d = 132.6\pm1.9$~pc.
Because of this ``long'' distance and correspondingly high implied luminosity,
Feast \& Catchpole (1997) and van Leeuwen et al.\ (2007) concluded that Polaris
is a first-overtone pulsator. 

However, TKUG13 argue that the parallax of Polaris is considerably larger,
$10.10\pm0.20$~mas ($d=99\pm2$~pc). The evidence cited by TKUG13 for this
``short'' distance includes (1)~a photometric parallax for Polaris~B based on
measured photometry, spectral classification, and main-sequence fitting; (2)~a
claim that there is a sparse cluster of A-, F-, and G-type stars within
$3^\circ$ of Polaris, with proper motions and radial velocities similar to that
of the Cepheid, for which the \Hipp\/ parallaxes combined with main-sequence
fitting give a distance of 99~pc; and (3)~a determination of the absolute visual
magnitude of Polaris based on line ratios in high-resolution spectra, calibrated
against supergiants with well-established luminosities. On the basis of the
short distance, and thus a fainter absolute magnitude, TKUG13 concluded that
Polaris is a fundamental-mode pulsator.

The angular diameter of Polaris has been measured interferometrically (Nordgren
et al.\ 2000; M\'erand et al.\ 2006). For the short distance, the radius implies
that Polaris pulsates in the fundamental mode, whereas the larger radius if the
long distance is adopted means that it pulsates in the first overtone (e.g.,
Bono et al.\ 2001; Neilson 2014).

In a critique of the TKUG13 paper, van~Leeuwen (2013, hereafter L13) defended
the \Hipp\/ parallax by presenting details of the solution, concluding that
``the \Hipp\/ data cannot in any way support'' the large parallax advocated by
TKUG13\null. Using \Hipp\/ data, L13 also questioned the reality of the sparse
cluster proposed by TKUG13, presenting evidence against it both from the color
vs.\ absolute-magnitude diagram for stars within $3^\circ$ of Polaris, and their
non-clustered distribution of proper motions. Lastly, L13 examined the absolute
magnitudes of nearly 400 stars of spectral type F3~V in the \Hipp\/ catalog with
parallax errors of less than 10\%, and showed that the absolute magnitude of
Polaris~B would fall well within the observed $M_V$ distribution for F3~V stars,
based on either the \Hipp\/ parallax of A or the larger parallax proposed by
TKUG13\null. Thus, he concluded, the photometric parallax of B does not give a
useful discriminant.

Neilson (2014) has given an extended discussion of the astrophysical issues
related to the distance of Polaris, including a consideration of the measured
rate of change of the pulsation period. He concluded that the properties of
Polaris are inconsistent with it being in the early evolutionary stage of the
first crossing of the Cepheid instability strip. Instead, Neilson argued that it
must be in the third crossing. This would require it to be more luminous, its
distance to be at least 118~pc (parallax less than $\sim$8.5~mas), and it to be
pulsating in the first overtone. However, Fadeyev (2015), based on hydrodynamic
pulsation models, reached the opposite conclusion: Polaris is crossing the
instability strip for the first time and is a fundamental-mode pulsator.

In this paper we present a measurement of the trigonometric parallax of the
Polaris system based on astrometric observations of the companion, Polaris~B,
with the Fine Guidance Sensors (FGSs) on \HST\null. After discussing the data
acquisition and analysis, we present the parallax result---which favors the long
distance or indeed an even larger distance than found by \Hipp. We conclude with
brief discussions of the astrophysical implications for the Cepheid, the
apparent peculiarities of Polaris~B, and the possibilities that Polaris actually
pulsates in the {\it second\/} overtone or that Polaris~B is not actually a
physical companion of A.

\section{{\em Hubble Space Telescope\/} Astrometry of Polaris~B}

\subsection{FGS Observations and Data Analysis}

As part of an astrometric program on the trigonometric parallaxes of overtone
Cepheids, we observed Polaris with the FGS system on \HST\null. The FGSs are a
set of three interferometers that, in addition to providing guiding control
during imaging or spectroscopic observations, can measure precise positions of a
target star and several surrounding astrometric reference stars with one FGS
while the other two guide the telescope. The FGS system has been shown capable
of yielding trigonometric parallaxes, in favorable cases, with better than
$\pm$0.2~mas precision (e.g., Benedict et al.\ 2007, hereafter B07; Soderblom et
al.\ 2005; Benedict et al.\ 2011, 2017; McArthur et al.\ 2011; Bond et al.\
2013). 

The Cepheid Polaris~A, at a mean brightness $\langle V \rangle = 1.982$ (Fernie
et al.\ 1995), is too bright to be observed with the FGS system. Because of the
strong evidence that Polaris~B is a physical companion at the same distance as
the Cepheid (see above), we chose it instead as our astrometric target. We made
FGS observations of Polaris~B during two \HST\/ visits at each of five epochs
between 2003 October and 2006 September (program numbers GO-9888, -10113, and
-10482; PI H.E.B.), at dates close to the biannual times of maximum parallax
factor. We used FGS1r for the measurements, in its wide-angle astrometric
POSITION mode. There was no sign of duplicity of B in the FGS acquisition data.
In addition to Polaris~B, we observed a network of ten faint background
reference stars lying within $\sim$$5'$ of the target. Of the ten reference
stars, two were rejected because of acquisition failures, faintness, binarity,
or interference from the diffraction spikes of Polaris~A, and we retained eight
(with magnitudes of $V=14.1$--16.5) for the final solution. They are listed in
Table~1.

{\bf

Our FGS astrometric solution procedure is outlined by Bond et al.\ (2013), and
described in detail by B07 and Nelan (2017). The first step is to correct the
positional measurements from the FGS for differential velocity aberration,
geometric distortion, thermally induced spacecraft drift, and telescope pointing
jitter. Because of refractive elements in the FGS optical train, an additional
adjustment based on the $B-V$ color of each star is applied. Moreover, as a
safety precaution due to its proximity to Polaris~A, Polaris~B itself was
observed with the F5ND neutral-density attenuator, while the much fainter
reference stars were observed only with the F583W filter element. Thus it was
necessary to apply ``cross-filter'' corrections to the positions of Polaris~B
relative to the reference stars; the corrections are slightly dependent on
location of the star in the FGS field. 

The adjusted measurements from all ten visits were then combined using a
six-parameter overlapping-plate technique that solves simultaneously for scale,
translation, rotation, and proper motion and parallax of each star. Full
details, including the equations of condition, are given in B07, their section
4.1. We employed the least-squares program GAUSSFIT (Jefferys et al.\ 1988) for
this analysis.  Parallax factors are obtained from the JPL Earth orbit
predictor, version DE405 (Standish 1990). Since the FGS measurements provide
only the relative positions of the stars, the model requires input estimated
values of the reference-star proper motions and parallaxes, in order to
determine an absolute parallax of the target. These estimates (\S2.2) were input
to the model as observations with errors, which permits the model to adjust
their parallaxes and proper motions (to within their specified errors) to find a
global solution that minimizes the resulting $\chi^2$.

}

\subsection{Reference-Star Proper Motions and Parallaxes}

The initial proper-motion estimates for the reference stars were taken from the
UCAC5 catalog (Zacharias et al.\ 2017). In order to estimate the distances to
the reference stars, we employed spectral classification and photometry, and as
a lower-weight criterion, their reduced proper motions. For spectral
classification, we obtained digital spectra with the WIYN 3.5m telescope and
Hydra multi-object spectrograph at Kitt Peak National Observatory (KPNO), on the
night of 2003 November~22. The classifications were accomplished through
comparison with a network of MK standard stars obtained with the same
spectrograph, assisted by equivalent-width measurements of lines sensitive to
temperature and luminosity. The results are given in the sixth column in
Table~1.

Photometry of the reference stars in the Johnson-Kron-Cousins {\it BVI\/} system
was obtained at KPNO on one photometric night in 2007 October (0.9m telescope),
and on three photometric nights in 2008 October (2.1m telescope). Each star was
measured on between 9 and 13 individual CCD frames. The photometry was
calibrated to the standard-star network of Landolt (1992), and the results are
presented in Table~1. The internal errors of the photometry, tabulated in
Table~1, are generally quite small, but the systematic errors are probably
larger because of (a)~the high airmass at which the Polaris field has to be
observed, and (b)~the presence of a very bright star at the center of the field,
giving rise to PSF wings, diffraction spikes, and charge-bleeding columns across
much of the field.

Although Polaris itself is unreddened (e.g., Fernie 1990; Laney \& Caldwell
2007), or very lightly reddened [e.g., Gauthier \& Fernie 1978 find
$E(B-V)=0.02\pm0.02$, and TKUG13 give $E(B-V)=0.02\pm0.01$], it is known to lie
just in front of a molecular cloud, the ``Polaris Cirrus Cloud'' or  ``Polaris
Flare'' (e.g., Sandage 1976; Heithausen \& Thaddeus 1990; Zagury et al.\ 1999;
Cambr{\'e}sy et al.\ 2001; Ward-Thompson et al.\ 2010; Panopoulou et al.\ 2016;
and references therein). Thus significant reddening of the reference stars is
expected. 

To estimate their reddening, we compared the observed $B-V$ color of each star
with the intrinsic $(B-V)_0$ color corresponding to its spectral type
(Schmidt-Kaler 1982), from which we calculated an average $E(B-V)=0.25$. We also
used the extinction map of Schlafly \& Finkbeiner (2011), as implemented at the
NASA/IPAC website\footnote{\url{\tt http://ned.ipac.caltech.edu}}, to determine
the  reddening in the direction beyond Polaris. The Schlafly \& Finkbeiner map
gives a range of reddening values across the field covered by the reference
stars of $E(B-V)=0.26$ to 0.30, which is the {\it total\/} reddening for a
hypothetical star at a very large distance. We adopted a reddening of
$E(B-V)=0.25$ for all of the reference stars, except for R10, the nearest one,
for which we used $E(B-V)=0.21$ based on its spectral type and observed $B-V$.

The distances to the reference stars were then estimated as follows: (1)~For the
four stars classified as dwarfs, we used a calibration of the visual absolute
magnitude, $M_V$, against $B-V$ and $V-I$ colors derived through polynomial fits
to a large sample of nearby main-sequence stars with accurate photometry and
\Hipp\/ or USNO parallaxes, which is described in more detail in Bond et al.\
(2013). This algorithm corrects for effects of metallicity. (2)~For the four
subgiants, we searched the \Hipp\/ data for all stars classified with the same
spectral types that had parallaxes greater than 15~mas, and calculated their
mean absolute magnitude for use in the distance estimate.  For the dwarfs, our
$M_V$ vs.\ {\it BVI\/} calibration reproduces the known absolute magnitudes of
the sample of nearby dwarfs with an rms scatter of 0.28~mag. The scatter in the
subgiant $M_V$ calibrators was larger, $\sim$0.8~mag.  Our final estimated input
parallaxes and their errors, based on the scatter in the $M_V$ calibrators, are
given in the last column of Table~1, along with the output parallaxes given by
the $\chi^2$ solution.

\subsection{Parallax and Proper Motion of Polaris B}

Our solution results in an absolute parallax of Polaris~B of $6.26\pm 0.24$~mas
($d=158\pm 6$~pc), as indicated at the bottom of Table~1. The uncertainty
includes contributions from residual errors in the geometric-distortion
calibration of the FGS, errors in \HST\/ pointing performance, and errors in the
raw stellar position measurements. The proper-motion components for Polaris~B
from the FGS solution are\footnote{We abbreviate $\mu_{\alpha}\cos\delta$ with
the symbol $\mu_{\alpha^*}$} $(\mu_{\alpha^*},\mu_\delta) = (41.1\pm 0.4,
-13.8 \pm 0.4)\,\rm mas\,yr^{-1}$. The absolute proper motion of Polaris~A
determined by \Hipp\/ is $(44.48\pm 0.11, -11.85 \pm 0.13)\,\rm mas\,yr^{-1}$
(van Leeuwen 2007), but this includes an offset due to orbital motion in the
close A--Ab pair during the relatively short astrometric mission. The long-term
proper motion of A in the FK5 system, corrected to the \Hipp\/ frame, is
$(41.50\pm 0.97, -16.73 \pm 0.75)\,\rm mas\,yr^{-1}$, according to Wielen et
al.\ (2000).  Since the uncertainties of the individual UCAC5 proper motions
used to establish the FGS reference frame are about 1--$5\,\rm mas\,yr^{-1}$
(Zacharias et al.\ 2017), the agreement with the FGS results is reasonable.

\subsection{The Discrepancy with {\em Hipparcos}}

Our result for the parallax of Polaris~B ($6.26\pm 0.24$~mas) is 1.28~mas
smaller than found by \Hipp\/ for Polaris~A ($7.54\pm0.11$~mas). Is it plausible
that the \Hipp\/ result could be in error by such a large amount? 

\Hipp\/ parallaxes have usually agreed with the results of {\it HST}/FGS
measurements, or of other parallax techniques, to within their respective errors
(e.g., Benedict et al.\ 2002; McArthur et al.\ 2011; Bond et al.\ 2013).
However, there have been a few notable exceptions: (1)~For the Pleiades cluster,
Melis et al.\ (2014) obtained a precise cluster parallax of $7.35\pm0.07$~mas
from very-long-baseline radio interferometry (VLBI) astrometry of four
radio-emitting cluster members. FGS parallaxes of three other Pleiades stars
gave an average absolute parallax of $7.43\pm0.17$ (random) $\pm0.20$
(systematic)~mas (Soderblom et al.\ 2005), in accord with the VLBI result.
However, van Leeuwen (2009), based on \Hipp\/ astrometry of over 50 Pleiads,
found a mean cluster parallax of $8.32\pm0.13$~mas, larger by 0.97~mas than the
VLBI result. (2)~Benedict et al.\ (2011) used FGS to measure a parallax of the
Type~II Cepheid $\kappa$~Pavonis of $5.57 \pm 0.28$~mas; the \Hipp\/ parallax of
$6.52\pm0.77$~mas is larger by a similar 0.95~mas (although this is of lower
statistical significance because of the relatively large \Hipp\/ uncertainty).
(3)~VandenBerg et al.\ (2014) used FGS to measure parallaxes of three halo
subgiants. For two of them, the results agreed very well with \Hipp, but for
HD~84937, the \Hipp\/ value of $13.74\pm0.78$~mas was larger by 1.50~mas than
the FGS measurement of $12.24\pm0.20$~mas. (4)~Zhang et al.\ (2017) used VLBI
astrometry to derive a parallax of $4.42\pm0.13$~mas for the semi-regular
variable RT~Virginis, for which the \Hipp\/ parallax is $7.38\pm0.84$~mas, or
2.96~mas larger.

In summary, there are indeed isolated examples of the \Hipp\/ parallax
measurement being shown to be anomalously too large.

\subsection{Possible Sources of Systematic Error in the FGS Parallax}

In this subsection, we comment on possible causes of a systematic error in our
FGS parallax measurement for Polaris~B, which could potentially explain the
discordance with the \Hipp\/ value for the Cepheid Polaris~A.

(1). Could our input estimated parallaxes of the reference stars be
systematically too low by $\sim$1.3~mas? Omitting the star R10, which is
unusually nearby, we find a mean estimated parallax of the other seven reference
stars of 0.89~mas. This agrees quite well with the value of 1.0~mas for the mean
parallax of field stars at $V=15$, at the Galactic latitude of Polaris,
recommended by van Altena et al.\ (1995, their Fig.~2) based on a statistical
model of Galactic structure. Increasing our reference-star parallaxes by a mean
of about 1.3~mas would give serious disagreement with the van Altena et al.\
model values. Moreover, it would require the reference stars to be
systematically about 1.9~mag fainter in absolute magnitude than in our
calibration, which appears astrophysically unlikely---it would require all of
the main-sequence stars to be extreme subdwarfs, in conflict with their spectral
types.

(2). Was our ground-based CCD photometry affected by the presence of the  bright
Polaris~A in the frames? The required sense to give agreement with \Hipp\/ would
be that the reference stars are actually systematically brighter than indicated
by our measurements. Here we have a check, because the FGS measurements provide
independent estimates of the $V$ magnitudes, based on the observed count rates
and an approximate absolute calibration. Setting aside R7 and R8, which are the
angularly closest of the reference stars to the very bright Polaris~A, we find
our measured FGS magnitudes are an average of only 0.09~mag brighter than the
ground-based $V$ magnitudes. Such an amount is likely consistent with
contamination of the FGS photometric measurements by background scattered light
from Polaris. (Background scattered light is not subtracted from the measured
counts in the FGS reductions.)

(3). Did scattered light or dark counts affect the FGS astrometry? The Polaris
astrometric field is unique among those measured with the \HST\/ FGS system,
because of the presence of the extremely bright Polaris~A near the center of the
field. In addition to the magnitude measurements noted in the previous
paragraph, we indeed see evidence of scattered light across the field. This
shows up as enhanced count rates detected as the instantaneous $5''\times5''$
FGS field of view is slewed across blank sky from one reference star to the
next. However, this background light is faint, incoherent with the light from
the FGS target stars, and displays no significant gradient over the $\sim$$1''$
scale length of FGS interferometric measurements. Thus, the background only
slightly reduces the amplitude of the interference fringes, without
significantly displacing the measured positions. This is the same effect that
dark counts from the photomultiplier tubes have on the fringe amplitude of faint
stars ($V\ga14.5$), but likewise without systematically affecting their measured
positions. To verify these conclusions, we conducted extensive tests whereby
each reference star, as well as pairs and triplets of reference stars, were
removed from the solution to reveal any unusually affected individual exposures.
Removing reference stars increased the errors in the parallax measurements but
did not systematically change the parallax of Polaris~B by more than 0.3~mas. We
therefore conclude that the FGS measurement of the Polaris~B parallax was not
significantly affected by the presence of Polaris~A.

(4). What evidence does \Gaia\/ provide? The recent first \Gaia\/ data release
(DR1; Gaia Collaboration et al.\ 2016a,b) provides an additional test of our
results. Positions of Polaris~B and the FGS reference stars were tabulated in
DR1, but none of them are contained in the {\it Tycho}-\Gaia\/ Astrometric
Solution (TGAS), and thus none have as yet a \Gaia-based parallax or proper
motion. (Polaris~A was also not included in DR1 or TGAS, as it is too bright for
the standard \Gaia\/ pipeline processing.) However, we used the epoch 2015.0
\Gaia\/ positions for the reference stars and Polaris~B to simulate an
additional FGS observation set, and then combined them with the rest of our
data. We found excellent agreement of the FGS astrometry with the \Gaia\/
catalog positions (to better than 1~mas), but resulting in an even slightly
smaller parallax for Polaris~B of $5.90 \pm 0.29$~mas. Since we note that DR1
flags the positions of Polaris~B and the reference stars as being based upon a
``Galactic Bayesian prior for parallax and proper motion relaxed by a factor of
ten,'' we decided not to include the \Gaia\/ measurement in our final solution.
Nonetheless, the excellent agreement of the FGS and \Gaia\/ DR1 astrometry
strengthens our conclusion that our measurements have not been contaminated by
the presence of Polaris~A.

\section{Discussion: Puzzles of the Polaris System}

\subsection{Does Polaris Pulsate in the Second Overtone?}

Assuming the FGS parallax of $6.26\pm 0.24$~mas, a mean apparent magnitude of 
$\langle V \rangle = 1.982$, and a reddening of $E(B-V)=0.01\pm0.01$ (see
\S\S2.1--2.2), we find the mean absolute magnitude\footnote{For reasons given
in, e.g., VandenBerg et al.\ (2014, their \S5), we have not applied a
Lutz-Kelker correction to the absolute magnitude. In any case, the correction to
the $M_V$ of Polaris, using the formulation of Hanson (1979), would be only
about $-0.015$~mag, much smaller than the uncertainty of the value.} of
Polaris~A to be $\langle M_V \rangle = -4.07\pm0.09$. In Figure~1 we plot (black
filled circles) the Leavitt period-luminosity relation ($\langle M_V \rangle$
vs.\ logarithm of the fundamental pulsation period) for the following Galactic
Cepheids with well-determined distances: (1)~those for which B07 measured
trigonometric parallaxes with FGS; (2)~SY~Aurigae and SS~Canis Majoris, for
which parallaxes have been measured with \HST\/ spatial scans by Riess et al.\
(2014) and Casertano et al.\ (2016), respectively; (3)~the long-period Cepheid
RS~Puppis, for which the distance was determined from light echoes in the
surrounding dust (Kervella et al.\ 2014). The three red filled circles in
Figure~1 show the positions of Polaris under the assumptions that it pulsates in
the fundamental mode (marked ``F''), first overtone (``1O''), or second overtone
(``2O''). For the first overtone, we ``fundamentalized'' the period using the
relation given for Galactic Cepheids by Alcock et al.\ (1995), based on beat
Cepheids pulsating in both the fundamental and first overtone: $P_{\rm1O}/P_{\rm
fund}=0.720- 0.027\log P_{\rm fund}$. For the second-overtone period, we adopted
the ratio $P_{\rm2O}/P_{\rm1O}=0.8007$ from Antonello et al.\ (1986), based on
their data on the double-mode (first and second overtones) Cepheid CO~Aurigae.

Figure 1 indicates that, assuming the FGS parallax of B to be correct and
applicable to A, Polaris~A is likely to be pulsating in the second overtone, in
order for it to agree with the Leavitt law based on well-determined parallaxes
of Galactic Cepheids. (The \Hipp\/ parallax of Polaris~A yields an $\langle M_V
\rangle$ of $-3.66$, consistent with either first- or second-overtone pulsation.
The large parallax advocated by TKUG13 gives $\langle M_V \rangle=-3.03$ and
implies fundamental-mode pulsation, as they have argued.)

The second-overtone pulsation suggested for Polaris by Figure~1 is surprising
and puzzling. The MACHO and OGLE surveys have identified a number of first-,
second-, and even third-overtone Cepheids in the Large and Small Magellanic
Clouds (e.g., Soszy{\'n}ski et al.\ 2015a, 2015b). These data indicate a trend
toward higher overtones with decreasing stellar metallicity. Theoretically this
can be explained by an increase in the temperature range of the blue loop in the
H-R diagram to higher temperatures at lower metallicity. This results in a
larger fraction of overtone pulsators, which are hotter than fundamental-mode
pulsators. However, the longest periods observed for first-overtone pulsators in
the Magellanic Clouds are about 6 to 6.5 days. The longest periods for
second-overtone pulsators are about 1.6 days, considerably shorter than the
3.969~day period of Polaris. Moreover, the higher Galactic metallicity should
result in shorter upper-limit periods for first-overtone pulsators, and even
shorter ones for the second overtone.   

In addition, the radius of Polaris~A can be inferred from the angular diameter
given by interferometry ($3.123\pm0.008$~mas; M\'erand et al.\ 2006). For our
FGS-based distance, this implies a radius of $53.6\, R_\odot$.  At this radius,
the (non-canonical) period-radius relation of Bono et al.\ (2001) implies better
agreement with a first-overtone than a second-overtone pulsator.  

On the other hand, our suggestion that Polaris pulsates in the second overtone
appears to be in accord with some known properties of overtone pulsators. In
fact, Polaris has a number of characteristics which are uncommon in
fundamental-mode Cepheids, and might indicate pulsation in a mode beyond the
normal fundamental and first overtone:

(1). Polaris has an unusually rapid rate of period change, much faster than
would be expected for evolution through the instability strip (e.g., Neilson et
al.\ 2012). Overtone pulsators are known to have larger fluctuations and
instabilities in their pulsation cycles than fundamental-mode pulsators (e.g.,
Evans et al.\ 2002, 2015). This leads to the suggestion that the observed period
changes in overtone pulsators may be only partly driven by evolution through the
instability strip, and partly caused by instability in the pulsation cycles.
Evans et al.\ (2002) argued this is a natural consequence of the different
envelope locations---deeper in the envelope for the fundamental mode---that
dominate pulsation growth rates.  This effect could plausibly be expected to be
even more pronounced in a second-overtone pulsator. 

(2). Polaris is also virtually unique in having shown a long-term decrease in
pulsation amplitude, followed in recent years by a partial recovery (e.g., Evans
et al.\ 2002; Turner et al.\ 2005; Bruntt et al.\ 2008; Neilson et al.\ 2016;
and references therein). The only other known Galactic Cepheid with a remotely
similar variable amplitude is V473~Lyrae, which Moln{\'a}r \& Szabados (2014)
and Moln{\'a}r et al.\ (2017) have argued is a second-overtone pulsator. A more
extensive discussion of these properties will be given in a paper currently in
preparation (N.~R. Evans et al.).

Second-overtone Cepheids pulsating at a single period are at best rare. On the
other hand, without additional information such as a distance,  they are not
easy to identify among Galactic Cepheids, so our sample may be seriously 
incomplete. This is particularly true for very small-amplitude variables, for
which Fourier light-curve parameters, typically used as mode diagnostics, are
not available.

\subsection{Peculiarities of Polaris B}

Figure 2 shows a color-magnitude diagram (CMD; absolute $V$ magnitude vs.\ $B-V$
color) for Polaris~A and B, using absolute magnitudes calculated by applying our
FGS parallax to both stars. We also plot isochrones for solar metallicity and
ages of 75~Myr and 2.1~Gyr, obtained from the MIST website\footnote{\tt 
http://waps.cfa.harvard.edu/MIST/interp\_isos.html} (Dotter 2016; Choi et al.\
2016). The 75~Myr age was chosen so that the isochrone would pass through
Polaris~A, assuming it to be on the third passage through the Cepheid
instability strip. The 2.1~Gyr isochrone passes through our point for Polaris~B.

Although the 75~Myr isochrone satisfactorily reproduces the position of
Polaris~A, it fails badly for Polaris~B\null. The latter's absolute magnitude,
for our parallax of $6.26\pm0.24$~mas, an apparent magnitude of $V=8.65\pm0.02$
(Evans et al.\ 2008 and references therein), and $E(B-V)=0.01$, is
$M_V=2.60\pm0.08$. The isochrone magnitude at the color of Polaris~B [taken to
be $(B-V)_0=0.41\pm0.02$] is 3.73, with an uncertainty of about $\pm$0.13~mag
due to the uncertainty in the $B-V$ color and the steepness of the main
sequence.

The 2.1-Gyr isochrone passing through Polaris~B has an age far too large to be
consistent with Polaris~A\null. This is true even if we adopt the \Hipp\/
parallax, which would give Polaris~B an absolute magnitude of $M_V=3.00$, still
too bright. The large parallax advocated by TKUG13 gives $M_V=3.64$, in good
agreement with the young isochrone---which was one of their arguments for the
large parallax---but the direct parallax measurements by \Hipp\/ and FGS are
both considerably smaller.

We consider three alternatives to explain the apparently discrepant ages of
Polaris~A and B that arise if we assume that the FGS parallax applies to both
stars: (1)~In spite of the strong evidence presented in the literature and
summarized in our \S1, B is actually not a physical, coeval companion of the
Cepheid, but instead an unrelated, slightly evolved F dwarf with an age of
$\sim$2.1~Gyr. This interpretation requires a set of extraordinary coincidences
in angular separation, metallicity, radial velocity, and proper motion. This
appears highly improbable, but it is not physically impossible. (2)~Polaris~B is
a physical companion with the same young age as the Cepheid, but it is unusually
luminous. One possibility might be that it is an unresolved (even by FGS)
binary, but the FGS parallax places it more than the maximum possible 0.75~mag
above the main sequence. This would seem to require that B is currently in a
transitory state of high luminosity, perhaps because of a recent stellar merger.
This scenario may in principle be testable, e.g., by searching for a rapid
rotation rate or variability, or for an infrared excess; but such tests are made
difficult by the presence of the nearby, very bright Cepheid. (3)~The system's
age is in fact given by Polaris~B, $\sim$2.1~Gyr, and it is Polaris~A that
appears anomalously young. In this picture, A could be descended from a blue
straggler which merged at some time in the past. However, the position of A in
the CMD of Figure~2 requires a mass of about $5.9\,M_\odot$. The mass of B from
its position on the 2.1~Gyr isochrone in Figure~2, is $\sim$$1.5\,M_\odot$. This
makes it difficult to understand how Polaris~A could be descended from a blue
straggler of more than about $3\,M_\odot$.\footnote{The 75~Myr isochrone plotted
in Figure~2 is for non-rotating stars. The effects of rotation and main-sequence
convective overshoot on the evolution of Cepheids and their progenitors have
been studied by many authors (e.g., Anderson et al.\ 2014 and references
therein). Inclusion of rotation in the evolutionary sequences could reduce the
implied mass of Polaris~A by $\sim$10--20\% (Anderson et al.), but a
blue-straggler scenario would still be inconsistent with the low mass of
Polaris~B.} We could speculate that Polaris~A might have merged very recently
and could still be temporarily overluminous, but there is no direct evidence for
this, such as a high rotational velocity for the star.

\subsection{Summary}

We have used the FGS system on \HST\/ to measure the trigonometric parallax of
Polaris~B\null. We find a parallax of $6.26\pm 0.24$~mas, which is 1.28~mas
smaller than found by the \Hipp\/ mission for the primary star, the Cepheid
Polaris~A\null. Under the assumption that the Cepheid is a physical companion of
B, our result implies a high luminosity and suggests it is pulsating in the
second overtone of its fundamental mode. However, the location of B in the HR
diagram indicates that it is an evolved star with an age of $\sim$2.1~Gyr. The
discrepancy with the young age of the Cepheid appears to suggest one of two
possibilities: (1)~Polaris~B is actually a background star that is physically
unrelated to A (in spite of the strong evidence that it is a true companion); or
(2)~one of the stars in the system is peculiar: either the system is young and B
is in a transitory state of high luminosity, or the system is old, and it is A
that appears anomalously young.  It should be noted that even if the \Hipp\/
parallax of A is correct, these puzzles still exist as long as B is considered
to be a physical companion. These issues may be resolved once \Gaia\/ parallaxes
of both stars are available.

\acknowledgments

Support was provided by NASA through grants from the Space Telescope Science
Institute, which is operated by the Association of Universities for Research in
Astronomy, Inc., under NASA contract NAS5-26555. Support was also
provided by the Chandra X-ray Center
under NASA Contract NAS8-03060.
Based in part on observations at Kitt Peak National Observatory, National
Optical Astronomy Observatory, which is operated by the
Association of Universities for Research in Astronomy (AURA) under cooperative
agreement with the National Science Foundation. The authors are honored to be
permitted to conduct astronomical research on Iolkam Du'ag (Kitt Peak), a
mountain with particular significance to the Tohono O'odham.  
STScI summer students Ryan Leaman and Mihkel Kama assisted in data reduction for
the reference stars. We thank B. E. McArthur for calibration of the FGS1
geometric distortion, and F. van Leeuwen for comments on the \Hipp\/ parallax of
Polaris. 
This work has made use of data from the European Space Agency (ESA)
mission {\it Gaia} (\url{https://www.cosmos.esa.int/gaia}), processed by
the {\it Gaia} Data Processing and Analysis Consortium (DPAC,
\url{https://www.cosmos.esa.int/web/gaia/dpac/consortium}). Funding
for the DPAC has been provided by national institutions, in particular
the institutions participating in the {\it Gaia} Multilateral Agreement.

{\it Facilities:} \facility{HST (FGS), KPNO:2.1m, KPNO:0.9m, WIYN, Gaia}


\begin{figure}
\begin{center}
\includegraphics[width=5.5in]{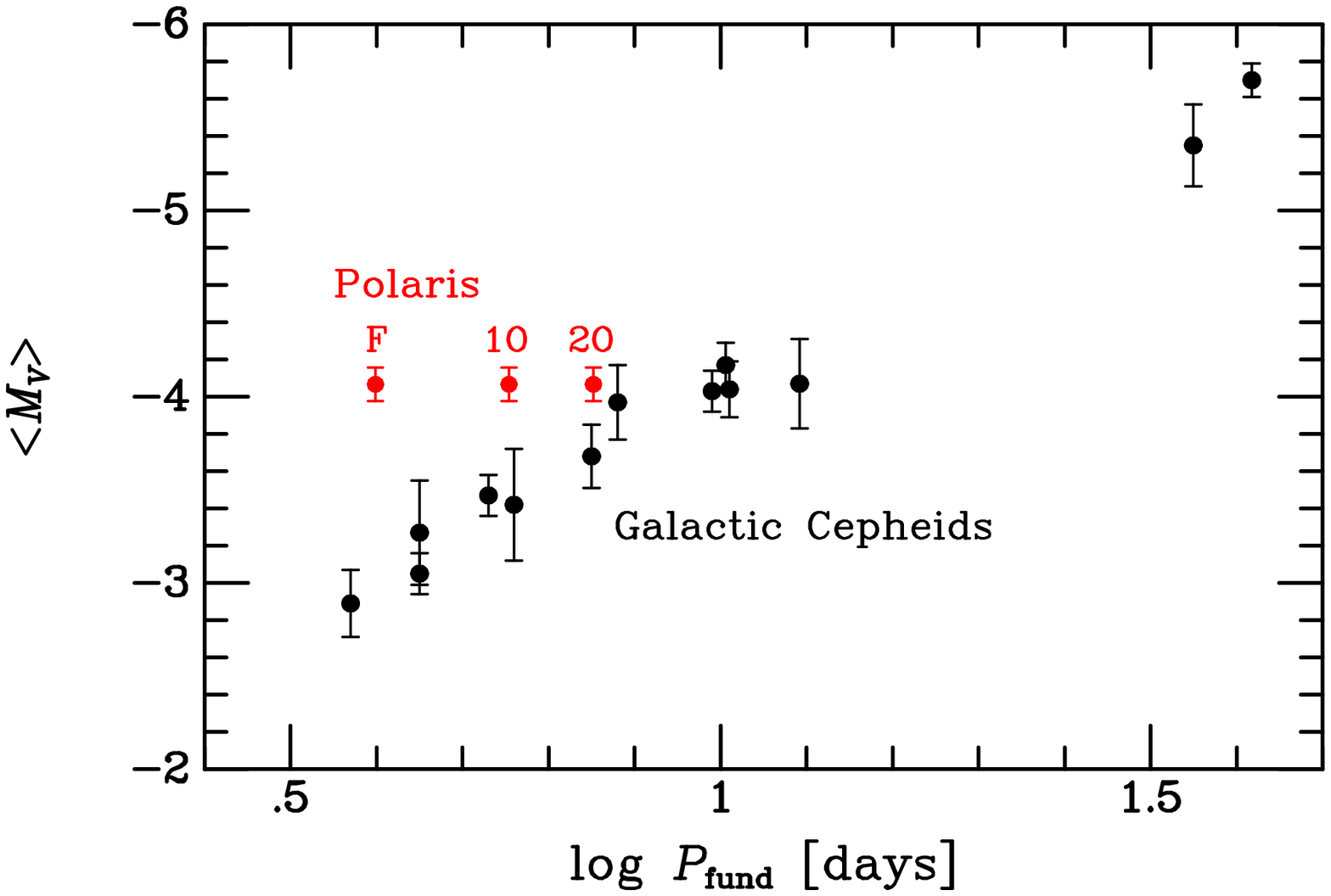}
\figcaption{
Location of Polaris in the Leavitt period-luminosity relation (mean absolute $V$
magnitude vs.\ logarithm of the fundamental-mode pulsation period), assuming the
FGS parallax of $6.26\pm 0.24$~mas measured for the companion star,
Polaris~B\null. The observed period is 3.969~days ($\log P=0.599$). If this is
the period of the fundamental mode, Polaris lies at the red filled circle marked
``F\null.'' If instead it pulsates in the first overtone, the corresponding
``fundamentalized'' period is the one marked ``1O\null.'' If the pulsation is at
the second overtone, the fundamentalized period is the one marked ``2O\null.''
The black filled circles show the Leavitt relation for Galactic Cepheids, based
on the FGS trigonometric parallaxes of Benedict et al.\ (2007), the \HST\/
spatial-scan parallaxes of Riess et al.\ (2014) and Casertano et al.\ (2016),
and the light-echo distance of RS~Puppis (Kervella et al.\ 2014).
}
\end{center}
\end{figure}

\begin{figure}
\begin{center}
\includegraphics[width=4.5in]{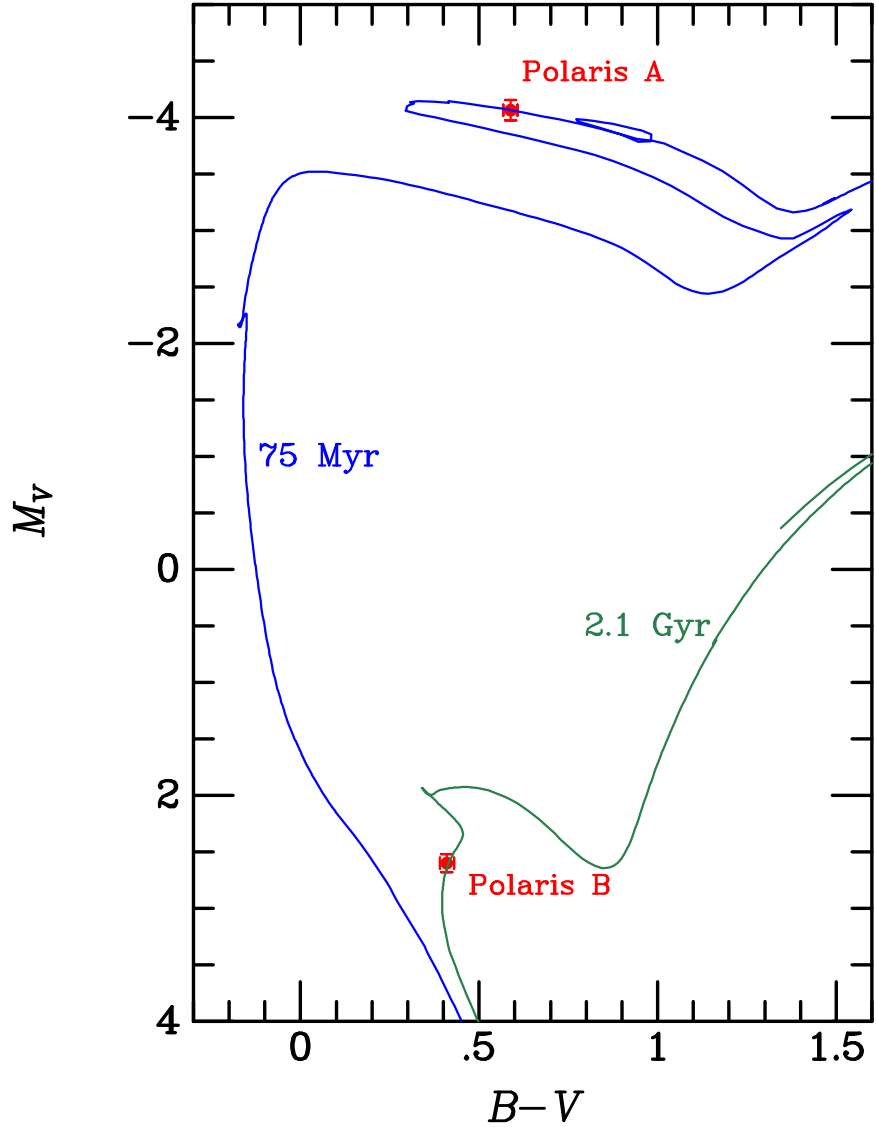}
\figcaption{
Locations of Polaris A and B (filled red circles) in the color-magnitude
diagram (absolute $V$ magnitude vs.\ $B-V$ color), based on the FGS parallax of
Polaris~B and the assumption that Polaris~A is a physical companion. Also shown
are solar-metallicity isochrones from the MIST data base (see text) for ages of
75 Myr (blue line) and 2.1~Gyr (green line).  
}
\end{center}
\end{figure}

\clearpage

\begin{deluxetable}{llccclcc}
\tabletypesize{\scriptsize}
\setlength{\tabcolsep}{0.05in}
\tablewidth{0pt}
\tablecaption{Astrometric Reference Stars and Polaris B}
\tablehead{
\colhead{ID} & \colhead{RA (J2000)} & \colhead{$V$} &
\colhead{$B-V$} & \colhead{$V-I$} & \colhead{Sp.Type} & 
\colhead{$\mu_{\alpha^*}$ [mas yr$^{-1}$]\tablenotemark{a}} & \colhead{$\pi_{\rm est}$ [mas]\tablenotemark{b}} \\
\colhead{} & \colhead{Dec (J2000)} & \colhead{$\sigma_V$} &
\colhead{$\sigma_{B-V}$} & \colhead{$\sigma_{V-I}$} & \colhead{} & 
\colhead{$\mu_{\delta}$ [mas yr$^{-1}$]\tablenotemark{a}} & \colhead{$\pi_{\rm final}$ [mas]\tablenotemark{b}} 
}
\startdata
R1  & 02:37:32.4 & 14.342 & 0.762 & 0.890 & F8 V & $0.9\pm0.4$ & $1.16\pm0.15$ \\
    & \llap{+}89:20:00.1  & $\pm$0.003 & $\pm$0.007 & $\pm$0.003 & & $-0.6\pm0.4$ & $1.14\pm0.07$ \\
R2  & 02:25:31.0 & 14.277 & 0.814 & 0.930 & G2 V & $-7.9\pm0.8$ & $1.31\pm0.17$ \\
    & \llap{+}89:18:09.5  & $\pm$0.003 & $\pm$0.004 & $\pm$0.004 & & $7.0 \pm0.5$ & $1.41\pm0.13$ \\
R3  & 02:34:04.9 & 16.504 & 0.734 & 0.820 & F7: IV: & $0.5 \pm0.8$ & $0.28\pm0.11$ \\
    & \llap{+}89:19:11.6  & $\pm$0.014 & $\pm$0.010 & $\pm$0.011 & & $-0.7\pm0.7$ & $0.28\pm0.04$ \\
R7\tablenotemark{c}  & 02:30:48.2 & 14.147 & 0.825 & $\dots$ & G0 IV & $5.4 \pm0.5$ & $1.04\pm0.35$ \\
    & \llap{+}89:14:30.2  & $\pm$0.003 & $\pm$0.007 & $\dots$ & & $0.5\pm0.4$ & $1.04\pm0.13$  \\
R8  & 02:25:26.6 & 15.304 & 1.116 & 1.237 & G0 IV & $9.7 \pm0.6$ &$0.49\pm0.16$ \\
    & \llap{+}89:14:26.2  & $\pm$0.015 & $\pm$0.011 & $\pm$0.009 & & $-6.8\pm0.5$ & $0.49\pm0.05$ \\
R9  & 02:21:18.2 & 14.958 & 0.903 & 1.070 & G1 IV & $13.3\pm1.0$ & $0.76\pm0.30$ \\
    & \llap{+}89:13:37.5  & $\pm$0.007 & $\pm$0.007 & $\pm$0.005 & & $1.5\pm0.7$ & $0.73\pm0.07$ \\
R10 & 02:32:25.8 & 14.675 & 1.360 & 1.633 & K5 V & $35.0\pm0.6$ & $5.48\pm0.70$ \\
    & \llap{+}89:12:09.2  & $\pm$0.004 & $\pm$0.008 & $\pm$0.007 & & $15.6\pm0.6$ & $6.32\pm0.42$ \\
R13 & 02:25:58.3 & 15.940 & 1.051 & 1.140 & G5: V: & $3.5\pm0.8$ & $1.16\pm0.15$ \\
    & \llap{+}89:12:12.9  & $\pm$0.006 & $\pm$0.020 & $\pm$0.010 & & $-2.0\pm0.7$ & $1.12\pm0.17$ \\
\noalign{\vskip0.1in}
B\tablenotemark{d} & 02:30:43.5 & 8.65   & 0.42  & $\dots$ & F3 V   & $41.1\pm0.4$ & $\dots$ \\
    & \llap{+}89:15:38.6  & $\pm$0.02  & $\dots$ &  $\dots$   & & $-13.8\pm0.4$ & $6.26\pm0.24$ \\
\enddata
\tablenotetext{a}{Proper motions in RA and Declination from our astrometric
solution}
\tablenotetext{b}{Input estimated absolute parallax (top entry), and adjusted
absolute parallax from astrometric solution (bottom entry)}
\tablenotetext{c}{R7 is cataloged as Polaris~D, which was identified as a
possible companion of Polaris by Burnham (1894), and discussed more recently by
Evans et al.\ (2002, 2010). The latter did not detect X-ray emission from
Polaris~D, suggesting that it is not a young low-mass companion of the Cepheid.
Our spectral type and photometry, giving an estimated distance of $\sim$960~pc,
and our measured proper motion, definitively rule out Polaris~D as a physical
companion of Polaris~A and~B.}
\tablenotetext{d}{Polaris B\null. $V$ magnitude from Evans et al.\ (2008) and
$B-V$ from literature compilation by Turner (2005); spectral type from Turner
(1977).}
\end{deluxetable} 


\end{document}